\documentclass{ws-procs9x6}

\def\adot{\dot{\alpha}}
\begin{document}

\title{Variation of the Fine-Structure Constant and Laser Cooling of Atomic Dysprosium}

\author{N. A. Leefer, A. Cing\"{o}z, D. Budker$^*$,}

\address{Department of Physics, University of California at Berkeley,\\
Berkeley, CA 94720-7300, USA\\
$^*$E-mail: budker@berkeley.edu}

\author{S. J. Ferrell, V. V. Yashchuk}

\address{Lawrence Berkeley National Laboratory,\\
Berkeley, California 94720, USA}

\author{A. Lapierre}

\address{TRIUMF National Laboratory,\\
Vancouver, British Columbia, V6T 2A3, Canada}

\author{A.-T Nguyen}

\address{Department of Otolaryngology, University of Pittsburgh,\\
Pittsburgh, Pennsylvania 15213, USA}

\author{S. K. Lamoreaux}

\address{Department of Physics, Yale University,\\
New Haven, Connecticut 06520-8120, USA}

\author{J. R. Torgerson}

\address{Physics Division, Los Alamos National Laboratory,\\
P-23, MS-H803, Los Alamos, New Mexico 87545, USA}

\begin{abstract}
Radio-frequency electric-dipole transitions between nearly degenerate,
opposite parity levels of atomic dysprosium (Dy) were monitored over
an eight-month period to search for a variation in the
fine-structure constant, $\alpha$. The data provide a rate of fractional temporal
variation of $\alpha$ of $(-2.4\pm2.3)\times10^{-15}$~yr$^{-1}$ or
a value of $(-7.8 \pm 5.9) \times 10^{-6}$ for $k_\alpha$, the
variation coefficient for $\alpha$ in a changing
gravitational potential.  All results indicate the absence of significant
variation at the present level of sensitivity.  We also present initial results on laser cooling of an atomic beam of dysprosium.
\end{abstract}

\bodymatter

\section{Introduction}\label{FSM:sec1}
A component of Einstein's equivalence principle (EEP)
is local position invariance (LPI), which states that the laws of
physics, including the values of fundamental constants, should be
independent of space and time. Modern theories attempting to unify
gravitation with the other fundamental interactions allow, or even
predict, violations of EEP~\cite{Damour2002}, which has sparked
searches for violation of LPI, and hence EEP, through searches for
temporal and spatial variation of fundamental constants.

Various studies have reported results of
searches for a temporal variation of the fine-structure constant
($\alpha = e^2 /\hbar c$) over cosmological time scales of $10^{10}$ years~\cite{Webb2001,Murphy2003,Quast2004,Srianand2004}, geological time scales of $10^9$ years~\cite{Damour1996,Fujii2000,Gould2006}, and present day laboratory searches over the course of years~\cite{Bize2003,Marion2003,Fischer2004,Peik2004,Fortier2007,Cingoz2007,Rosenband2008}.  In contrast to studies involving analyses of processes that have occurred billions of years ago, the results from laboratory searches are easier to interpret since the experiments are repeatable, and systematic uncertaintities can be studied by changing experimental conditions.  The best limit on a temporal variation that is independent of assumptions regarding other constants (published after our main result in Ref.~\refcite{Cingoz2007}) was obtained by monitoring the ratio of Al$^+$ and Hg$^+$ optical transition frequencies~\cite{Rosenband2008}.

Another type of search for an LPI violation is a ``null"
gravitational red-shift experiment where two clocks with different
composition are compared side by side in a changing gravitational
potential~\cite{Will}. Laboratory clock comparisons can be used for
this type of test due to the eccentricity of Earth's orbit around
the Sun, which leads to a small oscillatory component of the
gravitational potential with a period of a year. In earlier
work~\cite{Godone1994,Bauch2002,Fortier2007,Ashby2007}, clocks of
different types were compared, and the ratios of the clock rates
were analyzed for a possible correlation with the gravitational
potential, which led to bounds on parameters that characterize
structure-dependent modifications to the clock frequencies.  At the time of publication, we were able to present the best limit on $k_\alpha$, the linear-variation coefficient for $\alpha$ in a changing
gravitational potential~\cite{Ferrell2007}.  A better result has since been published in Ref.~\refcite{Blatt2008}.

Advances in direct comparison of single-ion optical or neutral
optical lattice clocks through frequency-comb metrology promise
significant improvements in sensitivity by up to three orders of
magnitude, as well as simplification in interpretation of the
results since such comparisons directly probe $\alpha$-variation,
independent of other fundamental constants (see, for example
Ref.~\refcite{Rosenband2008}). In this article, we present results obtained with
an alternative method of competitive sensitivity, which is also independent of other fundamental constants. This method utilizes rf electric-dipole transitions between nearly degenerate electronic levels in atomic Dy. 

\section{Effect of $\alpha$-variation in dysprosium}
\label{sec:1}

\begin{figure}[t]
\begin{center}
\psfig{file=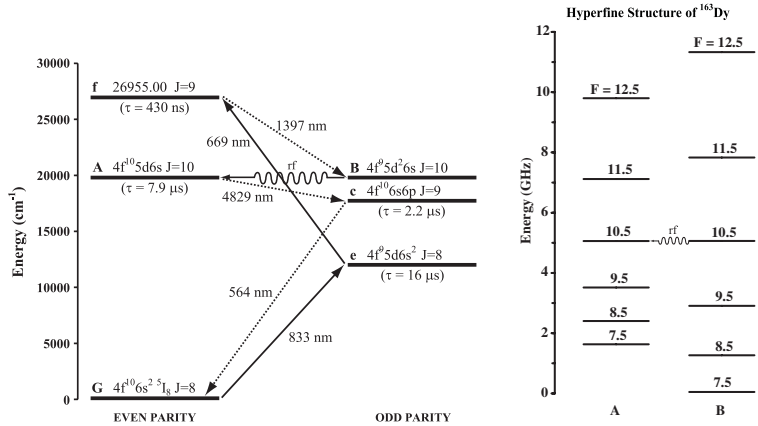,width=4.5in}
\end{center}
\caption{Relevant levels and transitions in
atomic dysprosium. Population and detection scheme: Level $B$ is
populated in a three-step process. The atoms are excited to level
$f$ in the first two steps using 833- and 699-nm laser light (solid
arrows). The third step is a spontaneous decay (labeled by a
short-dashed arrow) from level $f$ to level $B$ with $\sim 30$~\%
branching ratio. The population is transferred to level $A$ by the
rf electric field (wavy line). Atoms in level $A$ decay to level $c$
and then to the ground state. The fluorescence from the second decay
(564 nm) is used for detection. Inset on the right: Hyperfine
structure of levels A and B for $^{163}$Dy. Zero energy is
arbitrarily chosen to coincide with the lowest hyperfine component.}
\label{fig:popsch}
\end{figure}

Tests for variation of $\alpha$ in atomic systems rely upon the fact that the sensitivity of energy levels to $\alpha$ is different for different energy levels. The total energy of a level can be written as

\begin{equation}
\label{eq:levenergy}
    E = h\nu=E_0 + q(\alpha^2/\alpha_0^2 - 1),
\end{equation}
where $E_0$ is the present-day energy, $\alpha_0$ is the present-day
value of the fine-structure constant, and $q$ is a coefficient which
determines the sensitivity to variation of $\alpha$. A calculation~\cite{Dzuba2008}, utilizing relativistic
Hartree-Fock and configuration interaction methods, found values of $q$ for two nearly degenerate opposite-parity levels in atomic dysprosium (Fig.~\ref{fig:popsch}, levels $A$ and $B$) that are large and of opposite sign. For the even-parity level (level $A$), $q_A/hc\simeq8\times10^3$~cm$^{-1}$, while for the odd-parity level (level $B$), $q_B/hc\simeq-25\times10^3$~cm$^{-1}$.

\section{Experimental Technique}
\label{sec:2}
\begin{figure}[t]
\begin{center}
\psfig{file=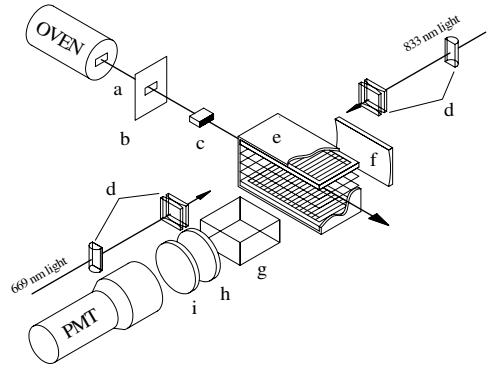,width=2.5in}
\end{center}
\caption{ Experimental setup (not shown to
scale): a) Atomic beam produced by effusive oven source at $\sim
1500$ K; b) atomic-beam collimator; c) oven light collimator; d)
cylindrical lenses to diverge laser beams; e) interaction region of
atoms with the electric field enclosed in a magnetic shield (not
shown); f) cylindrical mirror to collect fluorescent light; g)
lucite light pipe; h) interference filter; and i) short-pass
filter.}
\label{fig:oven&int}
\end{figure}

A unique aspect of the Dy system is that the directly observable
quantity is the energy difference of the levels, $\nu_B-\nu_A$, due to the fact that an
electric-dipole transition can be induced between them. The time variation of the transition frequency between levels $A$
and $B$ is related to temporal variation of $\alpha$ by

\begin{equation}
\label{eq:tempvar}
\dot{\Delta\nu}=\dot{\nu}_B-\dot{\nu}_A=2\frac{q_B-q_A}{h}\frac{\adot}{\alpha},
\end{equation}
where $|2(q_B-q_A)/h|\simeq 2.0 \times 10^{15}$~Hz~\cite{Dzuba2008}.
For instance, $|\adot/\alpha|=10^{-15}$/yr implies $|\dot{\Delta
\nu}|\simeq2$~Hz/yr.

A preliminary analysis of statistical and systematic
uncertainties shows that the measurement of the transition frequency
and the control of the systematics is feasible at a mHz level, which
corresponds to an ultimate sensitivity of $|\adot/\alpha|\sim10^{-18}$/yr for two measurements separated by a year's time~\cite{Nguyen2004}. A mHz resolution on a transition frequency of 1~GHz requires a relatively modest fractional stability of $10^{-12}$ for the reference frequency standard. This also means that the results are insensitive to variation of the Cs reference frequency due to changes in the values of fundamental constants, which have experimental upper limits$\sim 10^{-15}$~yr$^{-1}$ (see, for example, Ref.~\refcite{Marion2003}).

The transitions utilized in the experiment are the 235-MHz (J = 10
$\rightarrow$ J = 10) transition of the isotope $^{162}$Dy and the
3.1-MHz (F = 10.5 $\rightarrow$ F = 10.5) transition of $^{163}$Dy.
The long-lived level $B$ (lifetime~\cite{Budker1994} $\tau>200\ \mu$s) is populated via three transitions.  Lasers at 833 nm and 669 nm are used for the first two transitions to a high-lying upper state, which then spontaneously decays to level $B$ as shown in Fig.~\ref{fig:popsch}. The transition between levels $A$ and $B$ is induced by an applied rf electric field.  The atoms decay from level $A$ in two steps, and the fluorescence at 564 nm is monitored to detect the rf transition.

The rf-generation and detection system is discussed in Ref.~\refcite{Cingoz2005}. The rf field is frequency modulated by the 10 kHz reference output of a lock-in amplifier with a modulation index of 1. In order to reduce lineshape asymmetries due to drifts, the ratio of the first- and second-harmonic outputs of the lock-in amplifier is used to measure the transition frequency, which is extracted from the ratio by a two-step process described in detail in Ref.~\refcite{Cingoz2005}.

\section{Results and Analysis}
\label{sec:4}
\subsection{Temporal Variation}
\label{sec:4a}

\begin{figure}[t]
\begin{center}
\psfig{file=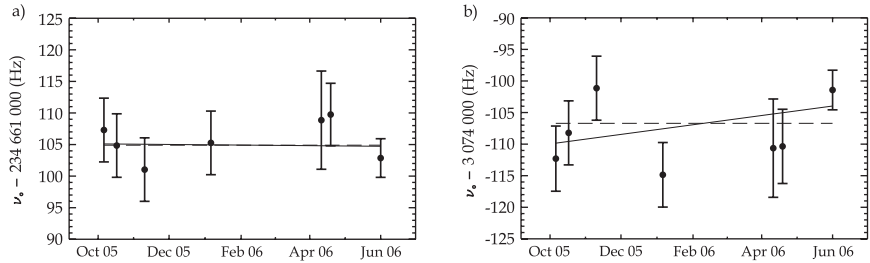,width=4.5in}
\end{center}
\caption{ Measured transition frequencies for
(a) the 235-MHz and (b) 3.1-MHz transitions over an eight-month
period. The data have been corrected for collisional shifts. The solid lines
are the least-squares linear fits to the data. The dashed lines are
the least-squares fits to a constant function. The apparent
correlation between the transition frequencies for the two isotopes
are due to the stray magnetic field effect.}
\label{fig:tempresult}

\end{figure}

We have measured the two rf transition frequencies in the course of seven runs over eight months. Figure~\ref{fig:tempresult} shows the results of these measurements corrected for collisional shifts~\cite{Cingoz2005}.  The present uncertainty is dominated by systematic effects primarily due to polarization imperfections coupled to the residual magnetic field, collisional shifts, and rf electric-field inhomogeneities.  A detailed analysis of these systematic effects and a discussion of feasible levels of stability are given in Ref.~\refcite{Nguyen2004}. Error bars are due both to statistical and systematic uncertainties which have been added in quadrature. A least-squares linear fit to the data points gives slopes of $-0.6\pm 6.5$~Hz/yr and $9.0\pm6.7$~Hz/yr for the 235-MHz and 3.1-MHz transitions, respectively.  Using an analysis presented in Ref.~\refcite{Cingoz2007}, we arrive at the final result of $\adot/\alpha=(-2.4\pm2.3)\times10^{-15}~\mbox{yr}^{-1}$, consistent (1$\sigma$) with no variation at the present level of sensitivity.  This result is slightly improved over what is presented in Ref.~\refcite{Cingoz2007} due to recently improved calculations of the sensitivity coefficients, $q_A$ and $q_B$~\cite{Dzuba2008}.

\subsection{Gravitational-Potential Dependence}

The data used in the previous analysis can also be analyzed for the dependence of $\alpha$ on variations of a gravitational potential.  The data were taken over the course of several months, and because the Earth's orbit has a slight eccentricity, each measurement took place at a different gravitational potential.

\begin{figure}[t]
\begin{center}
\psfig{file=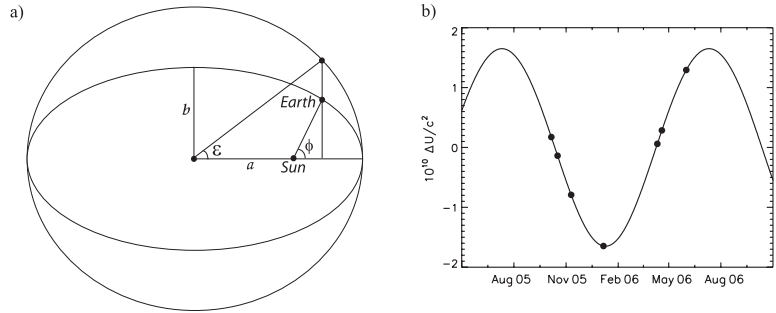,width=4in}
\end{center}
\caption{a) The true anomaly, $\phi$, is the angle subtended from
perihelion, the point of closest approach.  The eccentric anomaly,
$\varepsilon$, is the angle between perihelion and the position of
the Earth in its orbit projected onto the auxiliary circle of the
ellipse (the eccentricity of the ellipse is exaggerated for
clarity). b) The change in gravitational potential of the Sun
at the Earth due to the ellipticity of the orbit. The points indicate the dates on which data were taken.
\label{fig:grav}}
\end{figure}

The fractional change in $\alpha$ as a function of gravitational potential, $U(t)$, can be
parameterized as~\cite{Flambaum2007}
\begin{equation}
\label{eq:gravdep}
\frac{\delta\alpha}{\alpha} = k_{\alpha} \frac{\Delta U(t)}{c^2}\mbox{ ,} \\
\end{equation}
where $k_\alpha$ is the variation coefficient for $\alpha$ in a changing gravitational potential, $c$ is the speed of light, and $\Delta U(t)$ is the oscillatory component of the sun's gravitational potential at the Earth (see Fig.\ref{fig:grav} and Ref.~\refcite{Ferrell2007}).  The measured frequencies for each isotope are fitted to
the gravitational potential using a two-parameter least-squares fit
given by
\begin{equation}
\label{eq:fit}
 \delta(\Delta\nu) - \nu^*= x_0\frac{\Delta U(t)}{c^2}
+ x_1\mbox{ ,}
\end{equation}
where $\nu^*$ is an arbitrary reference frequency, and $x_0$ and
$x_1$ are the fit parameters. The parameter $x_1$ accounts for the
offset due to the reference frequency while the parameter $x_0$
determines the correlation between the change in gravitational
potential and the transition frequency.

\begin{figure}[t]
\begin{center}
\psfig{file=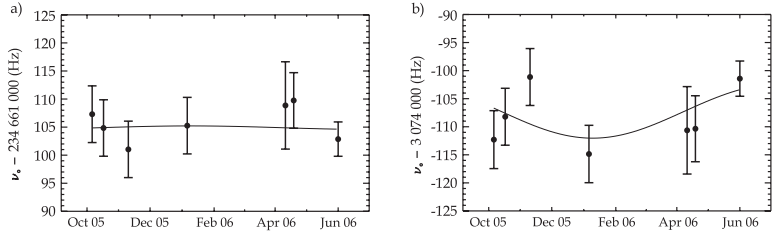,width=4in}
\end{center}
\caption{Same data as in Fig.~\ref{fig:tempresult}. Solid lines are
least-squares fit to a function given by Eq.\~ref{eq:fit} where
$\Delta U$ is the oscillatory component of the gravitational
potential of the Sun at Earth with a period of a year.
\label{fig:fits}}
\end{figure}

The parameters $x_0$ and $x_1$ were obtained by a least-squares fit to the data shown in Fig.~\ref{fig:fits}.  These values can be used to calculate the constraint on the parameter $k_{\alpha}$ from Eq.~(\ref{eq:gravdep}).  Substituting
Eq.~(\ref{eq:tempvar}) into Eq.~(\ref{eq:gravdep}), we get a
relation similar to Eq.~(\ref{eq:fit}),
\begin{equation}
\delta(\Delta\nu) = \left(2\,\frac{q_B - q_A}{h}\right)k_{\alpha}
\frac{\Delta U(t)}{c^2}\mbox{ .}
\end{equation}

The fit parameter yields $k_{\alpha} = (-1.0 \pm 8.2) \times 10^{-6}$ for the 235-MHz transition, $k_{\alpha} = (-14.6 \pm 8.4)
\times 10^{-6}$ for the 3.1-MHz transition. The final result of $k_{\alpha} = (-7.8 \pm 5.9) \times 10^{-6}$ is obtained from the analysis presented in Ref.~\refcite{Ferrell2007}.  In addition, our value of $k_{\alpha}$, combined with other experimental constraints, is used to extract the first limits on $k_e$ and $k_q$. These coefficients characterize the variation of $m_e/m_p$ and $m_q/m_p$ in a changing gravitational potential, where $m_e$, $m_p$, and $m_q$ are electron, proton, and quark masses. The results are $k_e = (4.3 \pm 3.5) \times 10^{-5}$ and $k_q = (5.9 \pm 4.7) \times 10^{-5}$.  As was noted in Sec.~\ref{sec:4a}, these results are improved over those presented in Ref.~\refcite{Ferrell2007} due to a more accurate calculation of sensitivity coefficients~\cite{Dzuba2008}.

\section{Laser Cooling of Dysprosium}

Recently, we have started to explore the possibility of laser cooling dysprosium, with the goal of generating a well collimated beam for our experiment.  This is expected to increase beam brightness, reduce sensitivity to spatial magnetic- and electric-field inhomogeneities, and allow us to characterize shifts associated with Dy-Dy collisions.

While dysprosium has been magnetically trapped before~\cite{Hancox2004}, there is no previously known effort to perform laser cooling. After a review of literature concerning excited state lifetimes~\cite{Curry1997,Wickliffe2000}, it was decided that the best option to attempt laser cooling was to use the 421.291-nm, $4f^{10}6s^2\,(J=8)\rightarrow4f^{10}6s6p\,(J=9)$ transition.  With an upper state lifetime of 4.8 ns, this transition is the strongest recorded cycling transition in the visible spectrum of Dy I.

\begin{figure}[t]
\begin{center}
\psfig{file=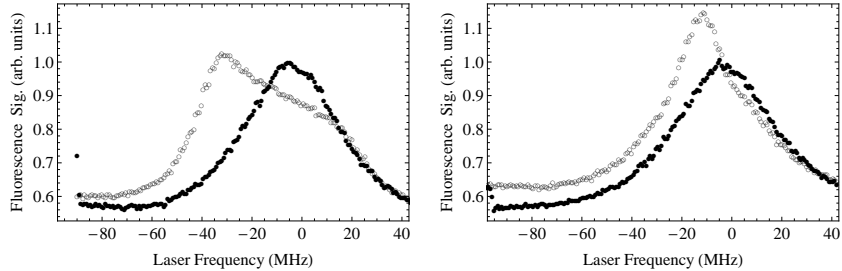,width=4.5in}
\end{center}
\caption{Scans of the $^{164}$Dy peak in the 658-nm transition with (hollow points) and without (solid points) the 421-nm pushing light incident on the atomic beam.  In the scan on the left the 421-nm light was tuned to the center of the $^{164}$Dy resonance, and in the scan on the right the 421-nm light was detuned from resonance by half the transverse doppler width.  In both scans an alteration of the lineshape, and hence velocity distribution, is clearly apparent.}
\label{fig:pushing}
\end{figure}

To generate the 421-nm light, 300 mW of 842-nm light from a Ti:Sapphire ring laser (Coherent 899) is frequency-doubled by a 1$\times$2$\times$10 mm$^3$ periodically poled potassium titanyl phosphate (PPKTP) crystal (Raicol Crystals Ltd.) located within a resonant bow-tie cavity.  The maximum output with this resonator-enhanced doubling scheme is approximately 90 mW.

In our initial attempt we choose to ``push" the atomic beam, as cooling required more changes to the experimental set-up.  To detect changes in the atomic beam, we use a parallel probe beam 1-cm downstream from the pushing beam.  Both the pushing and probing laser beams intersect the atomic beam at ninety degrees.  The 658.12-nm light for the probe laser is generated by a ring dye laser with DCM dye and resonant with the $4f^{10}6s^2 (J=8)\rightarrow4f^95d6s^2 (J=7)$ transition. The linewidth of this transition~\cite{Curry1997} ($\approx$100 kHz) is much less than the Doppler width (40 MHz), therefore the lineshape of the probe fluorescence is dominated by the transverse velocity distribution of the atomic beam.  A more detailed description of the atomic beam source is found in Ref.~\refcite{Nguyen1997}.

The results of the experiment are shown in Fig.~\ref{fig:pushing}.  For the first scan the 421-nm light is resonant with the center of the transverse velocity distribution, resulting in a shift and broadening of the probe lineshape.  In the second scan the 421-nm is red-detuned from the center of the velocity distribution, resulting in an increase in amplitude and narrowing of the lineshape.  In both scans the power in the pushing beam is approximately 60 mW in a 1.1-cm diameter, collimated beam.  Work is currently underway to modify the experiment for the purposes of more efficient transverse cooling.

Laser cooling of dysprosium is not trivial because of many available decay channels of the upper state of the cooling transition. However, our estimates (confirmed by more detailed atomic calculations~\cite{DzubaPriv2008}) show that the overall branching ratio into all excited states is $<10^{-3}$. In future work, we will determine the branching ratios and devise repumping schemes as needed.  In principle magneto-optical trapping, or even Bose-Einstein condensation may be possible. The latter would be of interest to the degenerate atomic gas community due to the large ground-state angular momentum and magnetic moment of Dy.

\section{Conclusion}
\label{sec:6}

We have presented improved results of a measurement of
time variation of$\alpha$ and a correlation with a changing gravitational
potential using atomic dysprosium.  In our case the interpretation does not require comparisons with different
measurements to eliminate dependence on other constants, and provides an alternative to measurements that utilize
state-of-the-art atomic optical frequency clocks. In addition, the data have been used to extract a limit on the gravitational variation of $\alpha$.  We have also presented results on the first effort to laser cool dysprosium.

A significant improvement in our results is expected from a newly operation apparatus, which provides better control over polarization imperfections coupled to the residual magnetic field, collisional shifts, and rf electric-field inhomogeneities, as well as other systematic effects expected to be important to achieve better than 1-Hz sensitivity. Ultimately, mHz-level sensitivity may be achievable with this method~\cite{Nguyen2004}.

\section{Acknowledgments}
We would like to thank V. V. Flambaum for intellectual inspiration
and collaboration on the analysis of the gravitational potential
dependence of $\alpha$ and V. A. Dzuba for calculation of matrix elements in Dy. We are grateful to D. F. Jackson Kimball,
and D. English for valuable discussions. This work was supported in
part by the University of California - Los Alamos National
Laboratory CLC program, NIST Precision Measurement Grant, Los Alamos
National Laboratory LDRD, and by grant RFP1-06-15 from the
Foundational Questions Institute (fqxi.org).

\bibliographystyle{ws-procs9x6}
\bibliography{alphadot}

\end{document}